\documentclass[conference]{IEEEtran}
\IEEEoverridecommandlockouts
\usepackage{cite}
\usepackage{amsmath,amssymb,amsfonts}
\usepackage{algorithmic}
\usepackage{graphicx}
\usepackage{textcomp}
\usepackage{float}
\usepackage{subcaption}
\usepackage{xcolor}
\def\BibTeX{{\rm B\kern-.05em{\sc i\kern-.025em b}\kern-.08em
    T\kern-.1667em\lower.7ex\hbox{E}\kern-.125emX}}
\usepackage{tikz}
\usetikzlibrary{positioning, arrows.meta, shapes.geometric}
\begin{document}

\title{Parallel Multi-Circuit Quantum Feature Fusion in Hybrid Quantum-Classical Convolutional Neural Networks for Breast Tumor Classification\\
}

\author{\IEEEauthorblockN{Ece Yurtseven}
\IEEEauthorblockA{\textit{Robert College of Istanbul} \\
Istanbul, Turkey \\
yurece.27@robcol.k12.tr}
}

\maketitle

\begin{abstract}
Quantum machine learning has emerged as a promising approach to improve feature extraction and classification tasks in high-dimensional data domains such as medical imaging. In this work, we present a hybrid Quantum-Classical Convolutional Neural Network (QCNN) architecture designed for the binary classification of the BreastMNIST dataset, a standardized benchmark for distinguishing between benign and malignant breast tumors. Our architecture integrates classical convolutional feature extraction with two distinct quantum circuits: an amplitude-encoding variational quantum circuit (VQC) and an angle-encoding VQC circuit with circular entanglement, both implemented on four qubits. These circuits generate quantum feature embeddings that are fused with classical features to form a joint feature space, which is subsequently processed by a fully connected classifier. To ensure fairness, the hybrid QCNN is parameter-matched against a baseline classical CNN, allowing us to isolate the contribution of quantum layers. Both models are trained under identical conditions using the Adam optimizer and binary cross-entropy loss. Experimental evaluation in five independent runs demonstrates that the hybrid QCNN achieves statistically significant improvements in classification accuracy compared to the classical CNN, as validated by a one-sided Wilcoxon signed rank test (p = 0.03125) and supported by large effect size of Cohen's d = 2.14. Our results indicate that hybrid QCNN architectures can leverage entanglement and quantum feature fusion to enhance medical image classification tasks. This work establishes a statistical validation framework for assessing hybrid quantum models in biomedical applications and highlights pathways for scaling to larger datasets and deployment on near-term quantum hardware.
\end{abstract}


\section{Introduction}
Breast cancer has emerged as a global health concern, affecting millions of women worldwide. In 2020, more than 2.3 million women were diagnosed with breast cancer, making it the most commonly diagnosed cancer globally \cite{arnold2022current}. Breast cancer remains a leading cause of cancer-related mortality, with an estimated 670,000 deaths recorded in 2022  \cite{kim2025global}. This burden is projected to continue rising in the following years. Early mammographic screening has been determined to play a critical role in lowering breast cancer mortality. Furthermore, AI-driven diagnostic systems have the potential to significantly improve the precision and efficiency of the screening process \cite{liu2023mammography}. However, the classical computational models that power these AI systems are beginning to encounter fundamental architectural and practical limitations, particularly in the demanding domain of medical imaging.

Classical Convolutional Neural Networks (CNNs) remain pivotal in modern computer vision, especially for biomedical image analysis, due to their efficacy in automatically extracting hierarchical features for diagnostic tasks like tumor detection \cite{mienye2025deep}. However, the ever-increasing complexity and volume of medical data often strain computational resources, challenging the scalability of classical deep learning algorithms. To address these limitations, Quantum Machine Learning (QML) offers a promising solution, leveraging quantum mechanics principles like superposition and entanglement for potential computational acceleration and robust pattern recognition \cite{biamonte2017quantum}.

Quantum computing represents a transformative approach with the potential to address the computational bottlenecks faced by classical machine learning \cite{gong2024quantum}. Using quantum mechanical properties such as superposition and entanglement, quantum computers offer the potential for exponential speedups in computation \cite{Nielsen_Chuang_2010}. This makes them suitable for complex computational tasks. Quantum algorithms can lead to several benefits over classical counterparts, offering exponential computational boosts and the ability to simultaneously process more inputs with fewer operations. In machine learning, quantum algorithms provide significant speedups in data handling by eﬀiciently managing high-dimensional vectors \cite{lloyd2013quantum}.

Building on these concepts, variational quantum circuits (VQCs) have become a widely used framework for near-term quantum machine learning. VQCs employ parameterized quantum circuits optimized through a hybrid quantum–classical loop, allowing models to access high-dimensional quantum representations while relying on classical algorithms for training \cite{cerezo2021variational}. This approach has supported applications such as quantum feature mapping, classification, and generative modeling, where VQCs can express complex transformations with relatively minimal circuits \cite{havlivcek2019supervised}. A key limitation is that single encoding schemes may not fully capture the complex features present in medical images, as different quantum encodings emphasize different aspects of classical data \cite{monnet2024understanding}.

Hybrid Quantum-Classical Convolutional Neural Networks (QCCNNs) have emerged as a particularly promising architecture that combines the strengths of both quantum and classical computing approaches \cite{liu2021hybrid}. These hybrid models utilize parameterized quantum circuits to construct quantum convolutional layers that perform linear unitary transformations on quantum states, extracting hidden information from encoded images, while classical fully connected layers process the measurement results for final classification \cite{liu2021hybrid}. This architecture not only leverages the powerful computing capabilities of quantum circuits but also inherits the established characteristics and training methodologies of classical neural networks. 

Hybrid quantum-classical convolutional neural networks have shown promising results for medical imaging applications. Studies demonstrate that parameterized quantum circuits can extract hidden information from medical images through linear unitary transformations which results in achieving faster training speeds and higher testing accuracy compared to classical approaches \cite{li2022image}. Quantum neural networks have emerged as a predominant approach for medical image analysis tasks such as cancer detection and diagnostic classification \cite{10859869}. These quantum-enhanced architectures offer computational advantages for processing high-dimensional medical imaging data. While various classical to quantum mapping methods ranging from basis encoding to amplitude encoding have been explored \cite{rath2024quantum}, the intricacies of quantum data encoding remain challenging, particularly the conversion of classical information such as images. 

In this work, we address these gaps by presenting a hybrid quantum-classical convolutional neural network architecture for the binary classification of breast tumors using the BreastMNIST dataset. Our key contributions are as follows: (1) introduction of a novel hybrid QCNN architecture using two parallel quantum circuits with amplitude and angle encoding schemes, (2) implementation of a quantum-classical feature fusion strategy that concatenates quantum features from both circuits with classical CNN features, (3) design of a parameter-matched experimental setup where hybrid and classical models share identical CNN backbones to isolate quantum contributions, and (4) demonstration of statistically significant performance improvements on a medical dataset validated across five independent runs.

\section{Related Works}

Xiang et al. \cite{xiang2024quantum} designed a QCCNN architecture for breast cancer diagnosis that combines classical CNN feature extraction with quantum convolutional layers. Their quantum layer initializes qubits in Hadamard ground states, applies parametric encoding through $U(\theta)$ gates with trainable rotation parameters, and performs Pauli-Z expectation measurements to extract quantum features for classification. The model demonstrated superior diagnostic performance over classical CNN across three breast cancer datasets: GBSG, SEER, and WDBC.

Matondo-Mvula and Elleithy \cite{matondo2024breast} introduced a Quanvolutional Neural Network architecture that embeds 3×3 image patches using angle encoding into a 9-qubit circuit, where each qubit is processed through a strongly entangled SU(4)-based kernel. Their model stacks two such quanvolutional layers as quantum convolutional filters, enabling the network to extract local features directly in Hilbert space before classification. Their model reached a peak validation accuracy of 87.17\% on the BreastMNIST dataset.

Tomal et al. \cite{tomal2024quantum} developed a hybrid QCNN using a 4-qubit quantum circuit. The sole quantum circuit consists of two back-to-back layers, one with angle embedding and one with circular entanglement, for the classification of the Iris dataset. Their approach integrates the quantum circuit as a feature extractor and then directly feeds its output into CNN for classification, demonstrating the effectiveness of their method by achieving 100\% classification \ accuracy for the Iris dataset.

Shahjalal et al. \cite{fahim2025hqcnn} proposed an HQCNN that integrates a five-layer classical CNN backbone with a 4-qubit variational quantum circuit that incorporates quantum state encoding, superpositional entanglement, and a Fourier-inspired quantum attention mechanism. Their model achieved 87.18\% accuracy on the binary classification of BreastMNIST dataset and 99.91\% accuracy for the PathMNIST dataset. Their hybrid quantum architecture employs cyclic U3 gate encoding with circular entanglement patterns to enable full qubit connectivity and quantum correlations.

Rifat et al. \cite{rifat2024enhanced} introduced a modified Dressed Quantum Network (DQN) where the conventional fully connected dimension reduction layer is replaced with a convolutional layer for feature extraction from ResNet50. Their hybrid classical-quantum architecture, tested on the BreastMNIST dataset, achieved an 69\% classification accuracy.

Kordzanganeh et al. \cite{kordzanganeh2023parallel} developed Parallel Hybrid Networks (PHN) that process inputs simultaneously through both a classical multi-layered perceptron and a variational quantum circuit, with outputs linearly combined using trainable weights to balance quantum and classical contributions. Their approach demonstrated that quantum circuits create smooth sinusoidal foundations while classical components fill non-harmonic gaps, achieving superior performance on periodic datasets with noise. 

Xu et al. \cite{xu2024parallel} proposed a Parallel Proportional Fusion model (PPF-QSNN) combining spiking neural networks with variational quantum circuits, where both networks process data in parallel and outputs are fused proportionally based on optimized quantum and classical contribution ratios. Their architecture achieved 97.1\% training accuracy on MNIST dataset by utilizing quantum superposition and entanglement alongside classical feature extraction.

Fan et al. \cite{10254235} introduced a hybrid quantum-classical CNN using amplitude encoding to represent images with 2n+1 qubits. Their quantum convolution layers use parameterized U3 gates for feature extraction with constant gate complexity. The model measures quantum states on multiple Pauli bases and feeds features to a classical dense layer for classification. Evaluated on five different datasets, it outperformed classical CNNs with fewer parameters and better generalizability.

\section{Methods}
\subsection{Dataset \& Preprocessing}
\begin{figure}[H]
    \centering
    \includegraphics[width=0.95\columnwidth]{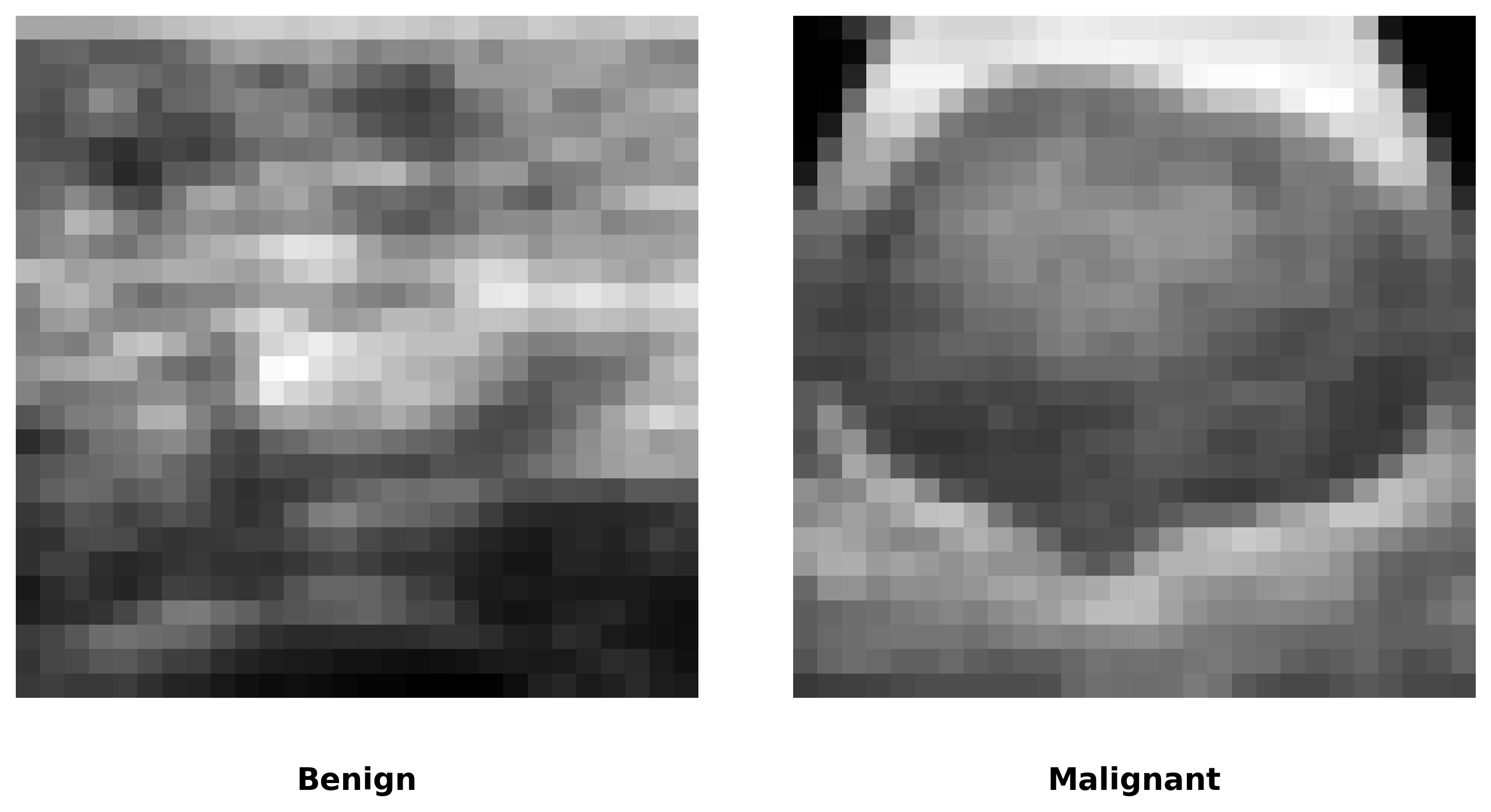}
    \caption{Representative training samples from the BreastMNIST dataset. Left: benign breast tissue. Right: malignant breast tissue.}
    \label{fig:breastmnist_samples}
\end{figure}

We utilize the BreastMNIST dataset from the MedMNIST collection \cite{medmnistv2}. The BreastMNIST dataset contains 780 standardized 28×28 pixel grayscale breast ultrasound images for the binary classification of benign and malignant tumors in breast tissue samples. Representative samples from the dataset are shown in Figure~\ref{fig:breastmnist_samples}. The dataset was split into training, validation and test sets following the standard BreastMNIST partitioning with 546 training, 78 validation and 156 testing samples. 

Additionally, in our experiments, all images were normalized from raw pixel values of [0, 255] to [-1, 1]. This normalization range is necessary to satisfy the input constraints of quantum circuits. For angle encoding, features are mapped to rotation angles via $R_Y(x_j \pi)$, requiring $x_j \in [-1, 1]$ to yield angles in $[-\pi, \pi]$ and amplitude encoding requires normalized feature vectors to represent valid quantum states.

\section{Variational Quantum Circuit Architectures}

\begin{figure}[htbp]
\centering
\includegraphics[width=0.5\textwidth]{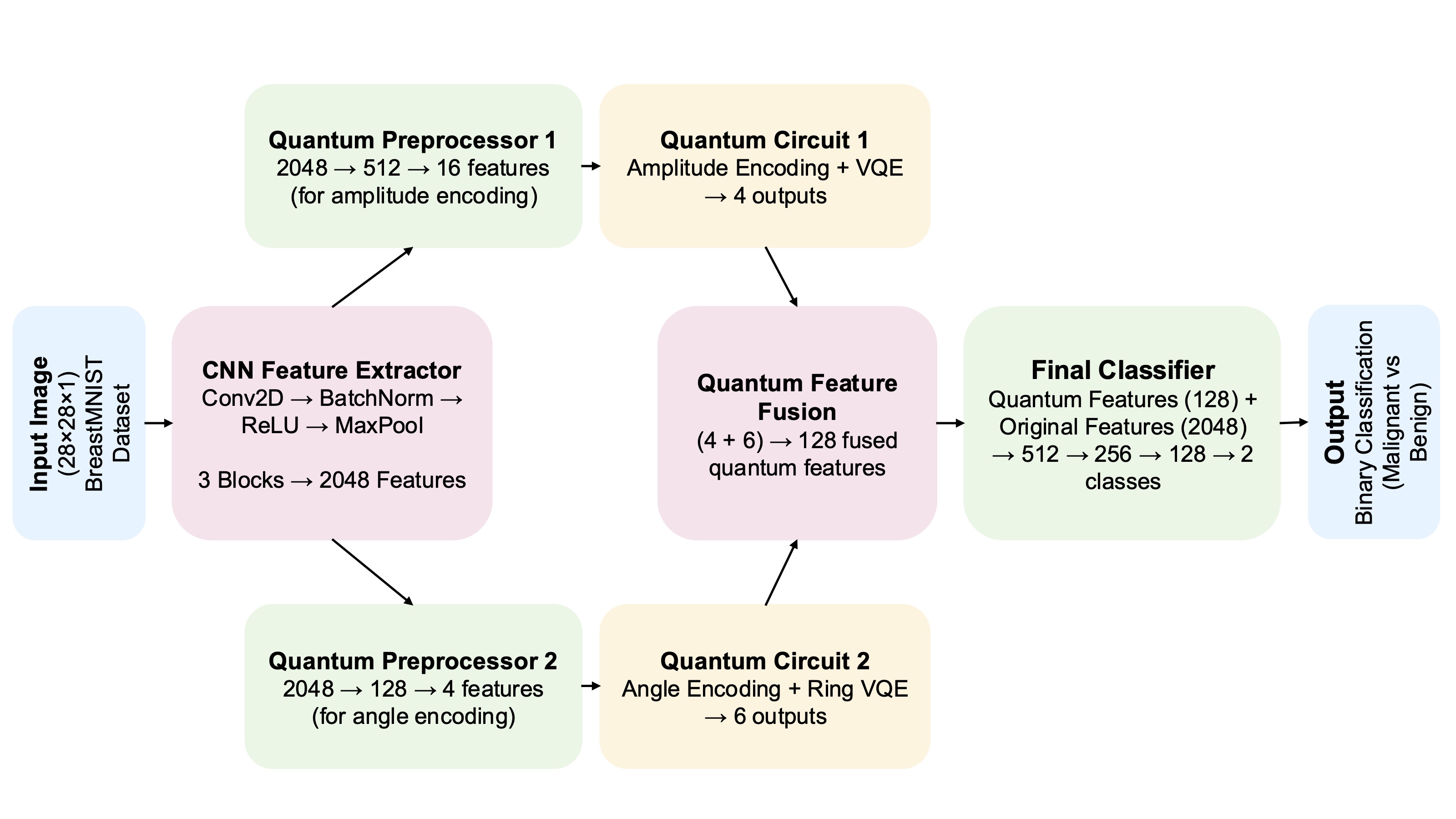}
\caption{Our proposed hybrid quantum-classical convolutional neural network architecture for breast cancer classification. Classical CNN features are processed through two parallel quantum pathways with different encoding schemes. 
}
\label{fig:architecture}
\end{figure}

Our hybrid QCNN utilizes two different variational quantum circuits (VQC), each implementing different data encoding strategies and entanglement structures to extract quantum features. Both circuits operate on 4 qubits and follow a layered architecture. In this section, we describe (A) the classical CNN backbone that extracts features from input images, (B) the amplitude encoding quantum circuit that embeds features into quantum state amplitudes, (C) the angle encoding quantum circuit with circular entanglement that maps features to rotation angles, (D) the quantum feature fusion mechanism that combines outputs from both quantum circuits with classical features, and (E) the training protocol used to optimize both hybrid and classical models.

\subsection{Classical Backbone (CNN Feature Extractor)}
Both quantum and classical models share an identical convolutional neural network feature extractor as their backbone architecture. This ensures a fair comparison by isolating the impact of quantum processing from classical feature learning. The feature extractor consists of six convolutional layers organized into three sequential blocks. Each block contains two Conv2D layers with batch normalization and ReLU activation, followed by pooling and dropout with rate 0.3 for regularization. The three blocks use 32, 64, and 128 filters respectively all with 3×3 kernels and padding 1. Max pooling operations progressively reduce spatial dimensions from 28×28 to 14×14 to 7×7 followed by adaptive average pooling to 4×4. The feature maps are then flattened to yield a 2048-dimensional feature vector of size 128 × 4 × 4. These extracted features, derived from the preprocessed images, serve as input to either quantum preprocessing layers or classical dense layers. The quantum preprocessors are implemented as two-layer fully connected 
networks. Preprocessor~1 performs a mapping from $2048$ to $512$ to $16$ 
features using \texttt{Linear} layers, combined with \texttt{LayerNorm}, 
\texttt{ReLU}, \texttt{Dropout}(p=0.3), and \texttt{Tanh} activation. 
Similarly, Preprocessor~2 maps $2048$ to $128$ to $4$ features with the 
same architectural components. The outputs are normalized to the interval 
$[-1, 1]$ to satisfy the input constraints of the quantum circuit.

\subsection{Quantum Circuit 1: Amplitude Encoding}

\begin{figure}[H]  
    \centering
    \includegraphics[width=\columnwidth]{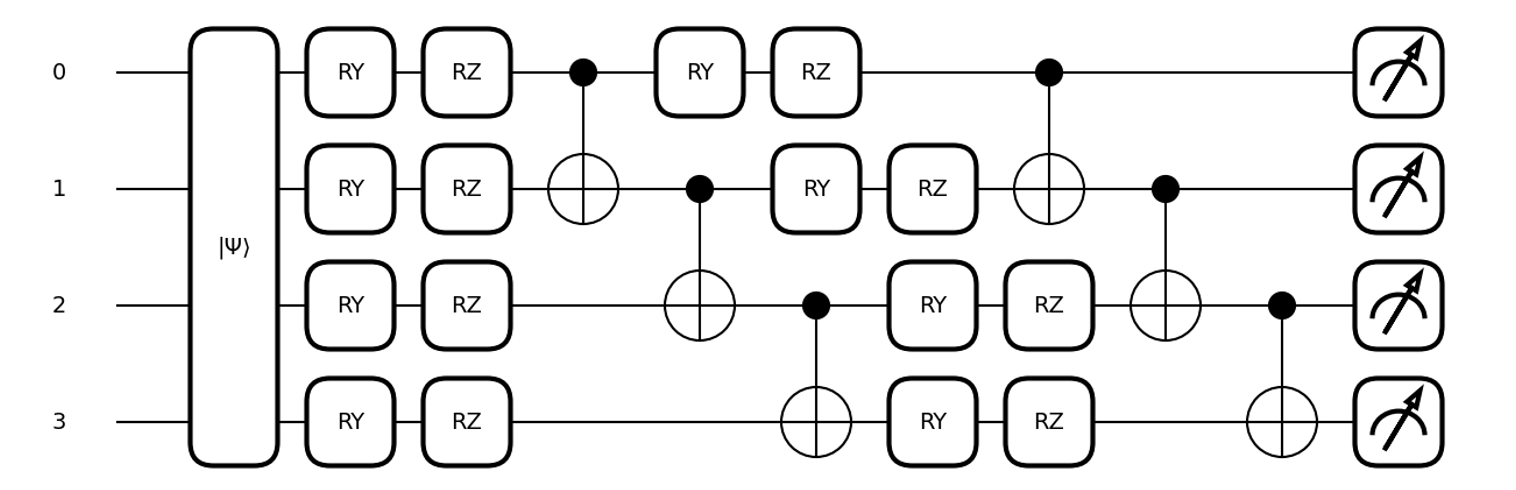}
    \caption{Amplitude embedding variational quantum circuit}
    \label{fig:amplitude_embedding_circuit}
\end{figure}

The first quantum circuit uses amplitude encoding to embed classical features into quantum state amplitudes:

\begin{equation}
|\psi\rangle = \sum_{i=0}^{15} \alpha_i |i\rangle, \quad \text{where } \alpha_i = \frac{f_i(x)}{\|f(x)\|_2}
\end{equation}

where $f(x) \in \mathbb{R}^{16}$ is the preprocessed feature vector and $\alpha_i$ are the normalized amplitude coefficients. This encoding efficiently maps 16 classical features into a 4-qubit quantum state coming from $2^4 = 16$ basis states. After variational quantum processing with trainable rotations and entanglement, we measure Pauli-Z expectation values on each qubit to obtain 4 quantum features, which can be represented as $\mathbf{q}_1 \in \mathbb{R}^4$.

\subsection{Quantum Circuit 2: Angle Encoding with Circular Entanglement}

\begin{figure}[H]
    \centering
    \includegraphics[width=\columnwidth]{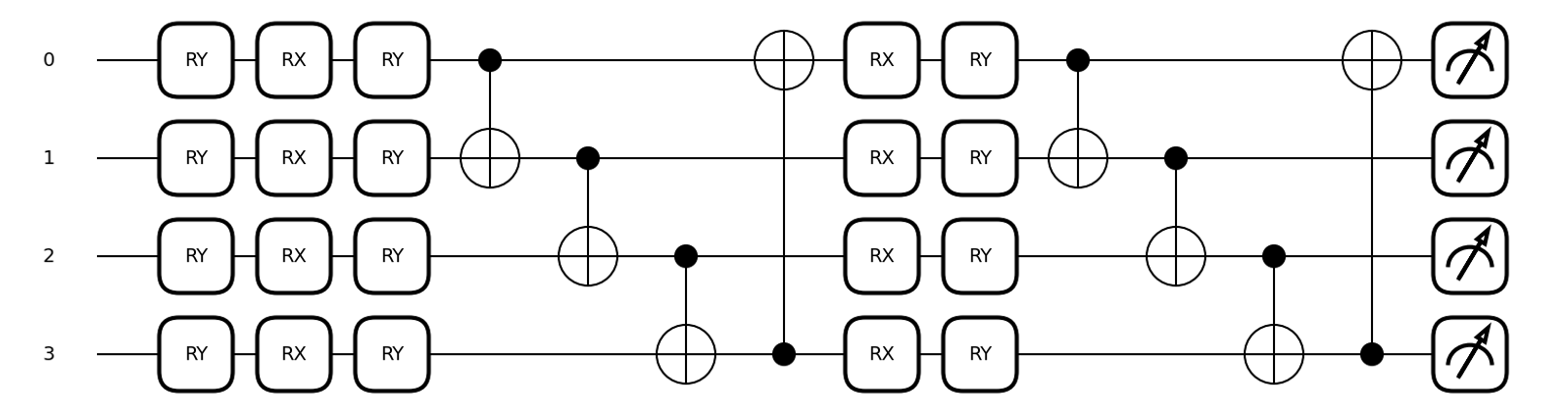}
    \caption{Angle encoding variational quantum circuit with circular entanglement}
    \label{fig:angle_encoding_circuit}
\end{figure}

The second quantum circuit employs angle encoding to map classical features directly into rotation angles:

\begin{equation}
R_Y(x_j \pi) |0\rangle, \quad j = 0, 1, 2, 3
\end{equation}

where $x_j \in [-1, 1]$ are the preprocessed feature values applied to each of the 4 qubits. Following the encoding layer, the circuit applies $L=2$ variational layers, each consisting of parameterized rotations $R_X(\theta_{l,j})$ and $R_Y(\phi_{l,j})$ with trainable parameters $\theta_{l,j}, \phi_{l,j} \in \mathbb{R}$.  Circular entanglement is implemented through CNOT gates according to the connectivity pattern:

\begin{equation}
\text{CNOT}_{j \to (j+1) \bmod 4}, \quad j = 0, 1, 2, 3
\end{equation}

This ring topology ensures that the final qubit connects back to the first, creating a closed entanglement loop that enables full qubit connectivity and quantum correlations throughout the circuit. Measurements are performed in Pauli-Z basis for all qubits and an additional Pauli-X basis for qubits 0 and 1, yielding 6 quantum features represented as $\mathbf{q}_2 \in \mathbb{R}^6$.

\subsection{Quantum Feature Fusion \& Classifier}
Following Gong et al.'s \cite{gong2024quantum} use of amplitude and angle encoding in VQC-based QCNNs, we extend their approach by implementing parallel multi-circuit quantum feature fusion rather than their tree-structured hybrid amplitude encoding method. 

Our proposed feature fusion stage concatenates two quantum feature vectors into a single 10-dimensional feature vector. Then the concatenated vector is passed through a learnable linear projection which is basically a weight matrix that maps the 10-dimensional input to a 128-dimensional space plus a learnable bias. The projected vector is then normalized using layer normalization, passed through a ReLU activation, and finally regularized with dropout. The output of this sequence is the fused 128-dimensional feature vector $h_{\text{fused}}$, which can be represented as:

\begin{equation}
\mathbf{h}_{\mathrm{fused}} = \mathrm{Dropout}\left(\mathrm{ReLU}\left(\mathrm{LayerNorm}\left(\mathbf{W} \cdot [\mathbf{q}_1; \mathbf{q}_2] + \mathbf{b}\right)\right)\right)
\end{equation}
where $\mathbf{q}_1 \in \mathbb{R}^4$ represents the output from the amplitude encoding quantum circuit, $\mathbf{q}_2 \in \mathbb{R}^6$ denotes the output from the angle encoding quantum circuit, and $[\mathbf{q}_1; \mathbf{q}_2] \in \mathbb{R}^{10}$ indicates their concatenation. The resulting fused feature vector $\mathbf{h}_{\mathrm{fused}} \in \mathbb{R}^{128}$ captures complementary information from both quantum encoding strategies.

The quantum-processed features are then integrated with the original classical CNN features to form a hybrid representation:

\begin{equation}
\mathbf{h}_{\mathrm{final}} = [\mathbf{h}_{\mathrm{fused}}; \mathbf{f}_{\mathrm{CNN}}]
\end{equation}
where $\mathbf{f}_{\mathrm{CNN}}$ denotes the classical features from the CNN backbone, resulting in $\mathbf{h}_{\mathrm{final}}$. This concatenation enables the model to leverage both quantum-enhanced representations and rich classical convolutional features.

The final classification is performed through a deep fully connected network:

\begin{equation}
\hat{y} = \mathrm{FC}(2176 \rightarrow 512 \rightarrow 256 \rightarrow 128 \rightarrow 2)
\end{equation}

The final layer outputs logits for binary classification, which are converted to class probabilities via softmax:

\begin{equation}
P(y = c | x) = \frac{\exp(\hat{y}_c)}{\sum_{j=1}^{2} \exp(\hat{y}_j)}
\end{equation}
where $c \in \{0, 1\}$ represents malignant and benign classes respectively.

\subsection{Training Protocol}
The classical baseline replaces each quantum circuit with a classical linear layer of identical input and output dimensionality. This preserves the same fusion architecture and classifier, ensuring parameter counts remain equivalent. Both models were trained using the AdamW optimizer with a learning rate of 0.001 and cross-entropy loss with label smoothing with a factor of 0.1. Quantum circuit weights were uniformly initialized in $[-\pi/6, \pi/6]$. A batch size of 16 was used, and early stopping with a patience of 25 epochs was implemented to prevent overfitting based on validation accuracy. Training was capped at 80 epochs, with the model checkpoint selected based on the highest validation accuracy. The OneCycleLR scheduler was employed with a maximum learning rate of 0.002 and cosine annealing strategy to dynamically adjust learning rates throughout training. Gradient clipping with a maximum norm of 1.0 was applied to stabilize training.

\section{Results}

\begin{table*}[t]
\centering
\caption{Comparison of Hybrid Quantum and Classical Model with Averaged Metrics}
\label{tab:classical_quantum_comparison}
\resizebox{\textwidth}{!}{%
\renewcommand{\arraystretch}{1.4}
\begin{tabular}{lcccccccc}
\hline
\textbf{Model} & 
\textbf{Training Accuracy} & 
\textbf{Validation Accuracy} & 
\textbf{Training Loss} & 
\textbf{Validation Loss} & 
\textbf{Testing Accuracy} & 
\textbf{Recall} & 
\textbf{Precision} & 
\textbf{F1-Score} \\
\hline
\textbf{Hybrid Quantum} & 
\textbf{0.8386} & 
\textbf{0.8795} & 
\textbf{0.4553} & 
\textbf{0.3786} & 
\textbf{0.8654} & 
0.9632 & 
\textbf{0.8670} & 
\textbf{0.9131} \\
Classical & 
0.8368 & 
0.8538 & 
0.4718 & 
0.3982 & 
0.8417 & 
\textbf{0.9649} & 
0.8397 & 
0.8967 \\
\hline
\end{tabular}%
}
\end{table*}

\subsection{Performance Metrics}

To compare the classical model and our proposed hybrid quantum model, the accuracy, recall, precision and F1 scores are used:

\begin{equation}
\text{accuracy} = \frac{TP + TN}{TP + TN + FP + FN}
\end{equation}

\begin{equation}
\text{recall} = \frac{TP}{TP + FN}
\end{equation}

\begin{equation}
\text{precision} = \frac{TP}{TP + FP}
\end{equation}

\begin{equation}
F_1 = \frac{2 \times \text{recall} \times \text{precision}}{\text{recall} + \text{precision}}
\end{equation}

\subsection{Training and Validation}
In the experiments, both models were trained for up to 80 epochs with early stopping applied depending on the validation accuracy. Training and validation accuracy were recorded with their corresponding loss at each epoch. The model checkpoint was selected based on the highest validation accuracy achieved during training. The experiments were repeated across 5 independent runs, per model, with random initializations to ensure statistical validity. Results from all 5 runs were aggregated to assess model performance consistency and in the tables, the values reported are depicting the average of 5 runs.

\begin{figure}[H]
    \centering
    \includegraphics[width=\columnwidth]{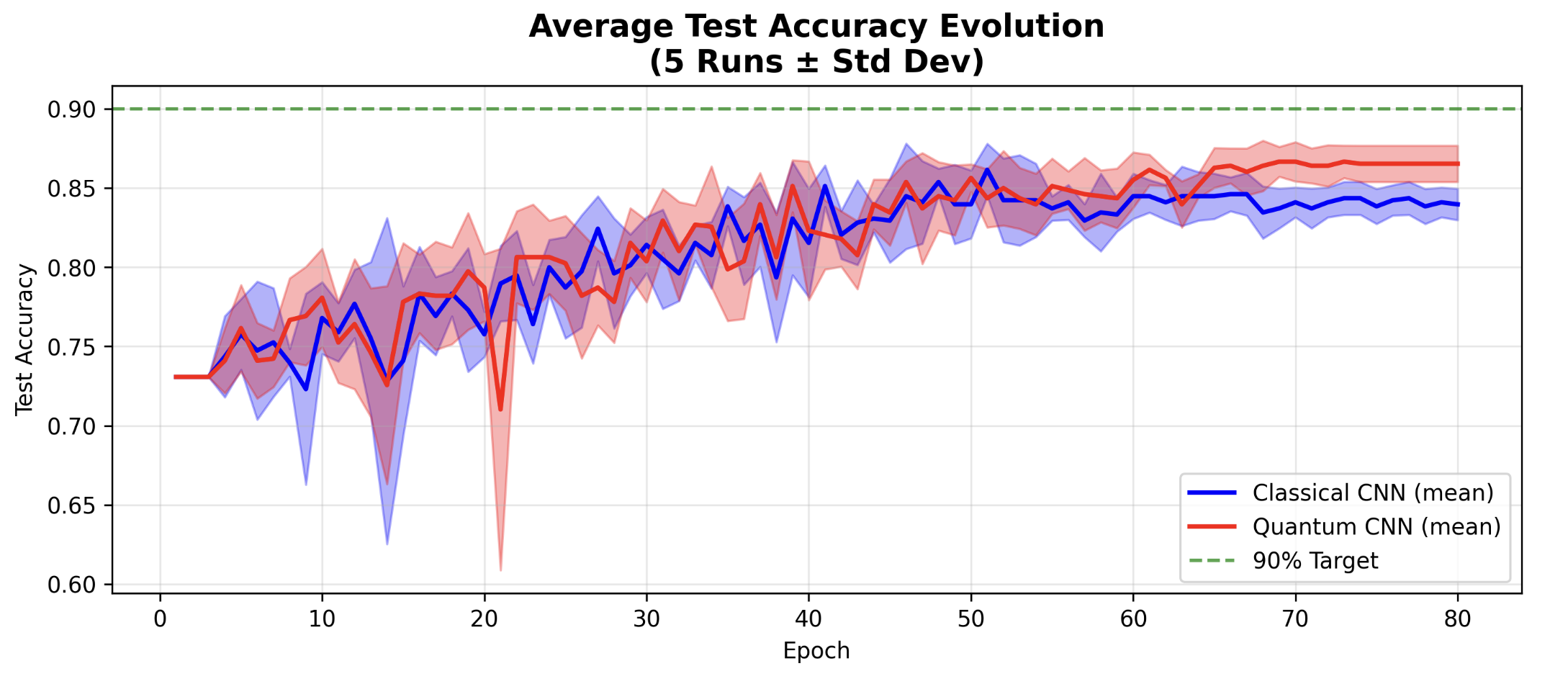}
    \caption{Testing accuracy graph including both models}
    \label{fig:train_accuracy}
\end{figure}

\subsection{Analysis of Metrics for Classical Model}
As shown in Figure 5, the classical CNN model achieved a final testing accuracy of 84.17\%. After 80 epochs, the model reported a training loss of 0.4718 and a validation loss of 0.3982, as seen in Figure 6.

\subsection{Analysis of Metrics for Hybrid Quantum Model}
After the same number of epochs, the quantum CNN had a training accuracy of 83.86\% with a loss value of 0.4553, and a validation accuracy of 87.95\% with a loss value of 0.3786 (Figure 7). As shown in Figure 5, the quantum model achieved a final testing accuracy of 86.5\%, significantly outperforming the classical model. 

\begin{figure}[H]
    \centering
    \includegraphics[width=0.65\columnwidth]{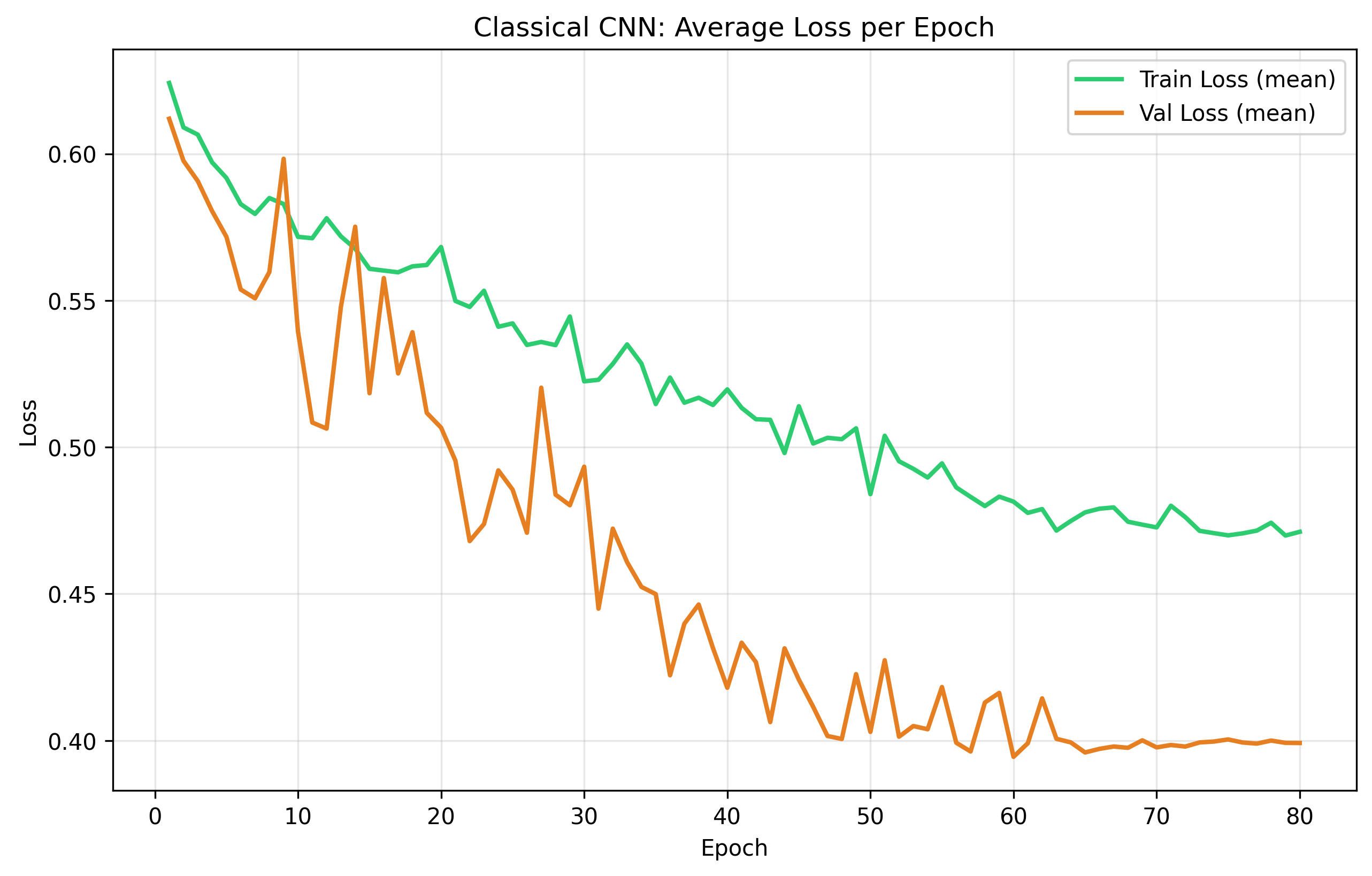}
    \caption{Training and validation loss curves for Classical Model}
    \label{fig:loss_classical}
\end{figure}

\begin{figure}[H]
    \centering
    \includegraphics[width=0.65\columnwidth]{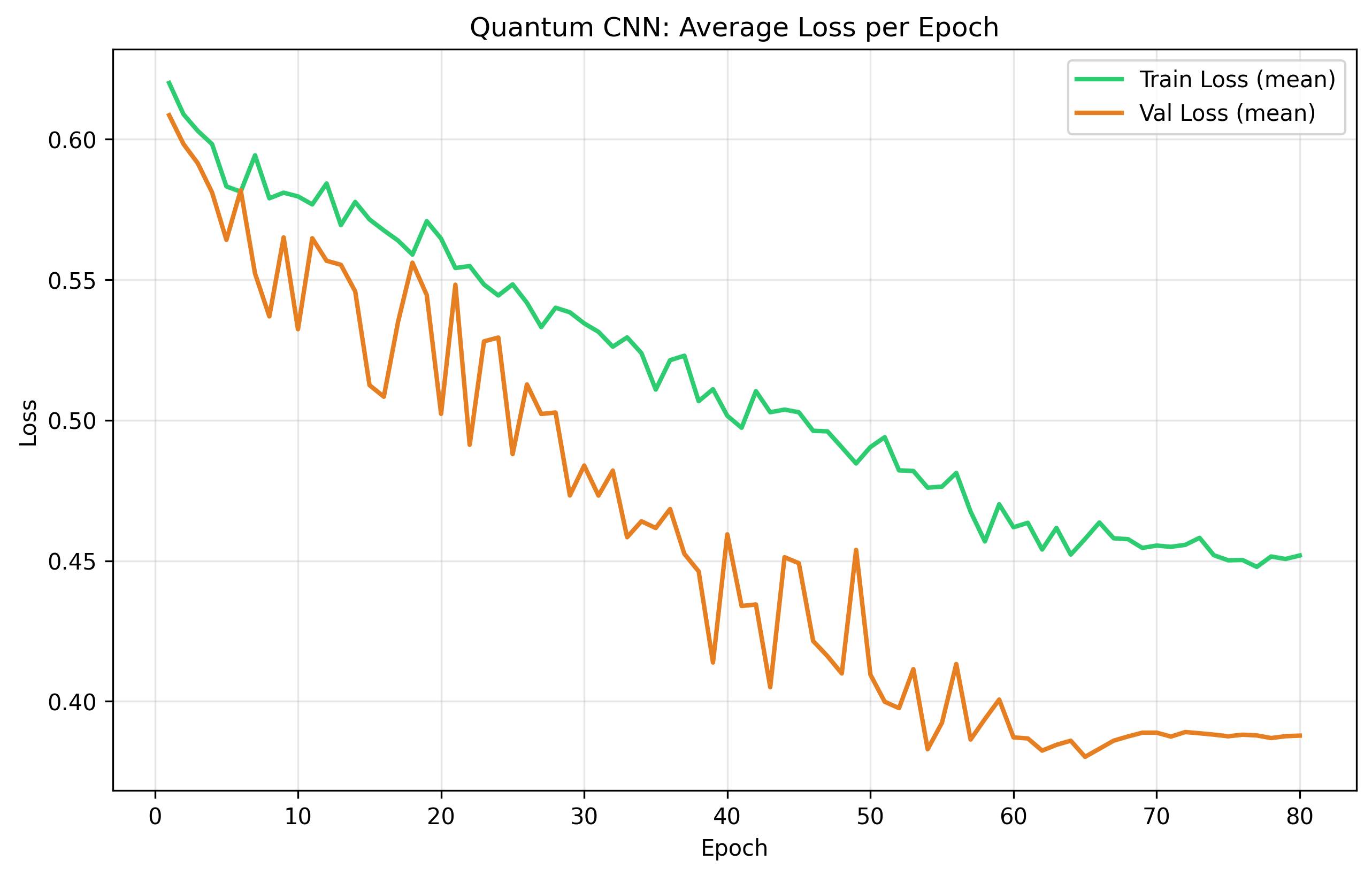}
    \caption{Training and validation loss curves for Hybrid Quantum Model}
    \label{fig:loss_quantum}
\end{figure}

\subsection{Testing}
The classical model testing results are shown in Figure 8b. Out of the 570 normal, benign images tested, 550 were correctly classified as benign and 20 were incorrectly classified as malignant. Additionally, out of the 210 malignant images tested, 105 were correctly classified as malignant and 105 were incorrectly classified as benign. In Figure 8a, which included the quantum CNN testing results, 126 malignant images were correctly classified as malignant and 84 malignant images were incorrectly classified as benign. The quantum model also correctly classified 549 benign images and incorrectly classified 21 benign images as malignant.

As shown in Table I, the quantum model performed better in the majority of metrics calculated: training accuracy, validation accuracy, training loss, validation loss, testing accuracy, precision, and F1-score. Since it has a higher accuracy and F1-score, it is better at correctly predicting the right label for an image across both classes, malignant and benign. Since the quantum model has a better precision than the classical model, it has a better accuracy for predicting positive labels out of the dataset. While the classical model has marginally higher recall, the quantum model demonstrates superior overall balance across performance metrics.

\begin{figure}[H]
    \centering

    \begin{subfigure}{0.49\columnwidth}
        \centering
        \includegraphics[width=\linewidth]{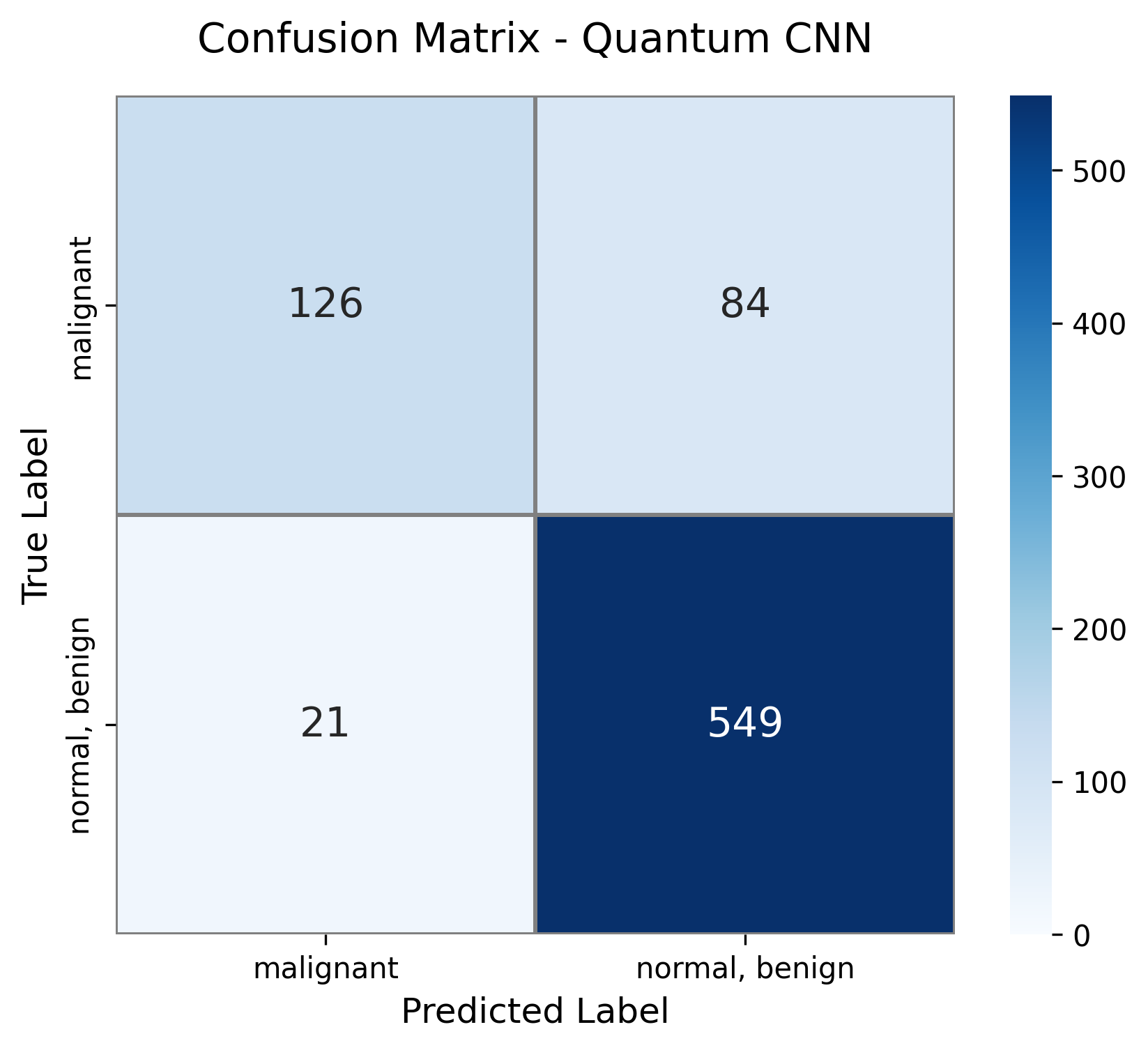}
        \caption{Hybrid Quantum}
        \label{fig:quantum_confusion_matrix}
    \end{subfigure}
    \hfill
    \begin{subfigure}{0.49\columnwidth}
        \centering
        \includegraphics[width=\linewidth]{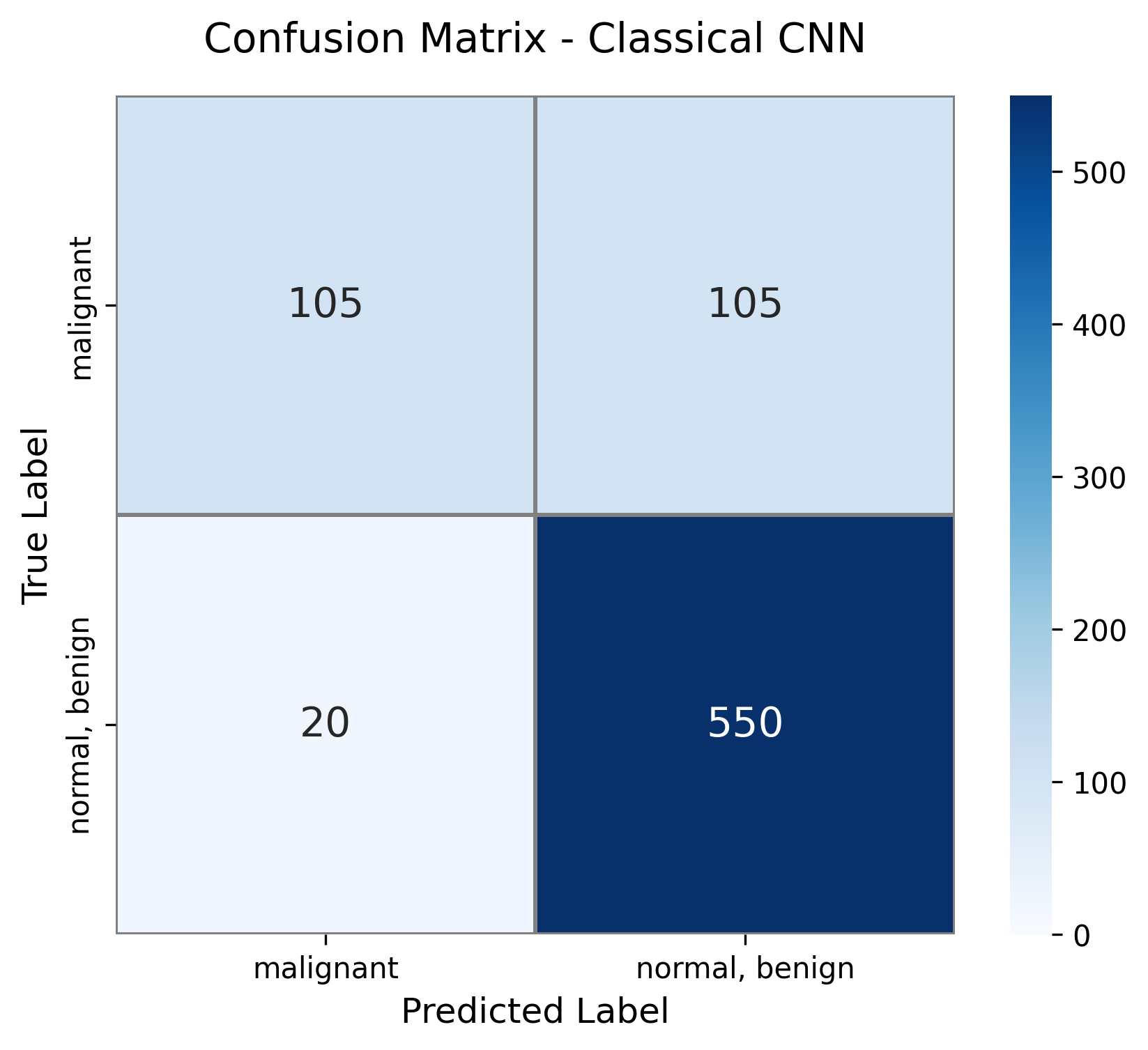}
        \caption{Classical}
        \label{fig:classical_confusion_matrix}
    \end{subfigure}

    \caption{Hybrid Quantum and Classical Confusion Matrices}
    \label{fig:confusion_matrices}
\end{figure}

\subsection{Statistical Validation Framework}

A one-sided Wilcoxon signed-rank test was conducted to compare the classification performance of the hybrid quantum CNN model and the classical CNN model across the five experimental runs. Following the approach of Fan et al. \cite{10254235}, who applied this test in a similar hybrid quantum-classical CNN context, we employed it here to assess performance differences between models. The null hypothesis states that the hybrid quantum model achieves an accuracy that is equal to or worse than that of the classical model, while the alternative hypothesis states that the hybrid quantum model achieves higher accuracy. The test yielded a p-value of 0.03125.

\[
p = 0.03125 < 0.05
\]

The resulting p-value ($p = 0.03125$) is below the 0.05 significance level, leading to the rejection of the null hypothesis in favor of the alternative. This provides statistical evidence that the hybrid quantum CNN model outperforms the classical CNN model under the evaluated conditions.

Furthermore, the effect size calculated using Cohen’s d was 2.14 which represents an exceptionally large effect. This indicates that the performance advantage of the hybrid quantum CNN over the classical CNN is practically substantial. In a similar study conducted by Ajlouni et al. \cite{ajlouni2023medical}, the authors also reported a comparably large Cohen’s d (d = 2.38) when evaluating performance differences between QCNNs and their classical counterparts, demonstrating that such large effect sizes are not unusual in this context.

\section{Discussion}
\subsection{Simulation Environment and Hardware Considerations}
All experiments in this study were conducted using quantum circuit simulations on classical hardware, which provides an idealized environment free from noise and decoherence effects inherent in current near-term intermediate-scale quantum (NISQ) devices \cite{preskill2018quantum}. All quantum expectation values were computed analytically using PennyLane's \texttt{default.qubit} statevector simulator with backpropagation. This corresponds to an idealized setting with infinite shots and no noise. In practice, performance on real quantum hardware may differ due to finite sampling and gate imperfections. While simulation allows for precise evaluation of the hybrid architecture's theoretical advantages, deployment on actual quantum hardware would introduce additional challenges including gate errors, limited qubit connectivity, and short coherence times. However, the relatively modest depth of our 4-qubit circuits and the use of simple gate operations (RX, RY, CNOT) make them suitable candidates for near-term hardware implementation. Future work should validate these results on real quantum processors to assess the practical viability of the proposed approach.

\subsection{Limitations}

Several limitations should be acknowledged for our proposed approach. First, our quantum circuits are restricted to 4 qubits due to simulation constraints which may potentially limit the expressiveness of quantum feature representations during the study. Also the BreastMNIST dataset, while standardized, is relatively small, which may limit the generalizability of our findings to larger and more diverse clinical datasets. Additionally, the dimensionality reduction required to map classical features into the quantum circuit's restricted qubit space may limit the expressiveness of feature representation.

\section{Conclusion}
In this work, we presented a hybrid quantum-classical convolutional neural network architecture for breast tumor classification that integrates two parallel variational quantum circuits with distinct encoding strategies. Our approach combines amplitude encoding and angle encoding with circular entanglement to generate complementary quantum feature representations, which are fused with classical CNN features for final classification. Experimental evaluation on the BreastMNIST dataset demonstrated that the hybrid QCNN achieves statistically significant improvements over a parameter-matched classical CNN baseline, with a testing accuracy of 86.54\% compared to 84.17\%. The statistical significance of these results was confirmed through a one-sided Wilcoxon signed-rank test (p = 0.03125) and with a large effect size of Cohen's d = 2.14. Our findings suggest that hybrid quantum-classical architectures can effectively leverage quantum mechanical properties such as superposition and entanglement to enhance medical image classification tasks. Future work should focus on scaling the quantum circuits, exploring alternative encoding methods, extending validation to larger clinical datasets, and implementing the model on actual quantum hardware with noise mitigation techniques.
\\

\bibliographystyle{IEEEtran}
\bibliography{IEEEabrv,MyBibFile}

\end{document}